\documentstyle[12pt]{article}

\def\be{\begin{equation}}
\def\ee{\end{equation}}

\newfont{\hbb}{msbm10 scaled \magstep1}
\def\setS{\mbox{\hbb S}}

\def\titel{Multiple Quantization and the Concept of Information}
\begin{document}
\sloppy

\thispagestyle{empty}

\begin{center}

\vspace*{5mm}

{\Large \titel}\footnote{Published:
{\em International Journal of Theoretical Physics}, Vol. 35, No. 11,
p. 2219 - 2225, 1996
}

\vspace*{5mm}

{\large Holger Lyre}\footnote{
Institute of Philosophy,
Ruhr-University Bochum,
D-44780 Bochum,
FRG,\\
email: holger.lyre@rz.ruhr-uni-bochum.de
}

\vspace*{5mm}

{\large May 1995}

\vspace*{10mm}

\begin{abstract}
The understanding of the meaning of quantization seems to be
the main problem in understanding quantum structures.
In this paper
first the difference between quantized particle vs. radiation fields
in the formalism of canonical quantization is discussed.
Next von Weizs\"acker's concept of ''multiple quantization''
which leads to an understanding of quantization as an iteration
of probability theory is explained.
Finally a connection between quantization and the idea of a
''general theory of information'' is considered.
This brings together semantic information with the different
levels of quantization and expresses the philosophical attitude
of this paper concerning the interpretation of quantum theory.
\end{abstract}

\end{center}

\vspace*{5mm}

\section{Quantum Field Theory}

When quantizing a field one has to differentiate between the
quantization of a classical radiation field such as the electromagnetic field
and a quantum field such as the Schr\"odinger, Klein-Gordon, or Dirac field.
Only in the latter cases is the field quantization indeed a ''second
quantization''. From a fundamental point of view the electromagnetic
field is one of the gauge fields in physics which describes one of the
fundamental forces, whereas the Dirac field describes the fundamental
fermions such as quarks and leptons. For the sake of simplicity I will only
consider the electron and the photon as examples of the fundamental particle
and gauge fields. Physically there is a clear difference between them:
the electron is a fermionic field of matter which describes particles
and the photon is a bosonic gauge field which describes interaction.

On the other hand the canonical formalism for the field quantization
seems to make no difference between this physical meaning of the
fields: they are both quantized fields
and therefore several authors maintain that there is no more
wave-particle-dualism on the level of quantum field theory.

According to the usual interpretation a quantized field is understood as
a totality of field quanta which can be created and annihilated.
Quantum field theory therefore is essentially a many-particle theory.

\subsection{The Dirac Field}\label{dirac_field}

The quantization of the Dirac spinors $\psi$, $\bar \psi$
leads to the operators
\begin{equation}\label{dirac}\begin{array}{rcl}
\hat \psi(x) &=& \sum_{\pm s} {\displaystyle \int}
   \frac{d^3p}{\sqrt{ (2\pi)^3 \frac{E}{m} }}
   \left( \hat b(p,s) u(p,s) e^{-ipx} + \hat d^+(p,s) v(p,s) e^{ipx}
   \right) \\
 & \\
\hat{\bar \psi}(x) &=& \sum_{\pm s} {\displaystyle \int}
   \frac{d^3p}{\sqrt{ (2\pi)^3 \frac{E}{m} }}
   \left( \hat b^+(p,s) \bar u(p,s) e^{ipx} + \hat d(p,s) \bar v(p,s)
   e^{-ipx} \right) \\
\end{array}\end{equation}
which describe the electron and the positron field,
i.e, particles and antiparticles.
The operators $\hat b^+(p,s)$, $\hat b(p,s)$, $\hat d^+(p,s)$ and
$\hat d(p,s)$ obey the commutation relations
\begin{equation}\label{diracCR}
\left\{\hat b(p,s), \hat b^+(p',s') \right\} =
\left\{\hat d(p,s), \hat d^+(p',s') \right\} =
\delta_{ss'} \delta^3(\vec p- \vec p')
\end{equation}
and zero otherwise.
Usually the canonical quantization procedure would lead to
Bose commutation relations instead of (\ref{diracCR}),
which for Dirac spinors violate microcausality.
Therefore anticommutation relations are needed which lead to fermions,
in agreement with experience.

Thus the Dirac field turns out to be an essentially complex-valued field,
i.e., the operators (\ref{dirac}) are non-Hermitian.
They do not describe quantities which are observed.
Measurable properties of the quantized Dirac field can only
be expressed in bilinear terms of the fields.
One therefore has to look at the operator of the probability density current
\begin{equation}
\hat \jmath^{\mu} = \hat{\bar \psi} \gamma^{\mu} \hat \psi
\end{equation}
which is conserved
\begin{equation}\label{continuity_eq}
\partial_{\mu} \hat \jmath^{\mu} = 0 .
\end{equation}

\subsection{The Electromagnetic Field}\label{EM_field}

In the case of the electromagnetic field the observed quantities are the
field forces $\vec E$ and $\vec B$ covariantly expressed by the tensor
\begin{equation} \label{fieldtensor}
F^{\mu\nu} = \partial^{\mu} A^{\nu} - \partial^{\nu} A^{\mu} .
\end{equation}
For the quantization of the electromagnetic field, however,
one should start from the potential $A^{\nu}$ because it appears
in the interaction terms and the transition amplitudes.
One then gets the operator
\begin{equation}\label{amu}
\hat A_{\mu}(x) = \int \frac{d^3k}{\sqrt{(2\pi)^3 2 k_0}}
\big( \hat a_{\mu}(\vec k) e^{-ikx} + \hat a_{\mu}^+(\vec k) e^{ikx} \big)
\end{equation}
where the Fourier ampitudes $\hat a_{\mu}(\vec k)$, $\hat a_{\mu}^+(\vec k)$
are to be regarded as photon annihilation and creation operators.
The canonical formalism leads to the commutation relations
\begin{equation}\label{emCR}
\left[\hat a(\vec k)  ,\hat a^+(\vec k')\right] = \delta(\vec k-\vec k')
\end{equation}
and zero otherwise.
In this paper I do not want to go into details concerning the special
problems of quantizing the electromagnetic field in a covariant manner
and to hold also only the two physical transversal polarization states
of the photon instead of the four degrees of freedom of the covariant
potential $A_{\mu}$.
These problems are related to the gauge freedom of $A_{\mu}$
and lead to the Gupta-Bleuler quantization.

Our interest is related to the question of whether the quantized
electromagnetic field can be regarded as a totality of photons
in the same manner as the Dirac field can be for electrons.
In this context two differences appear.
First, in contrast to the quantized Dirac field (\ref{dirac}),
the operator (\ref{amu}) is Hermitian.
This is an expression of the measurability of the quantized field
forces $\hat{\vec E}$ and $\hat{\vec B}$ - of course,
only within the scope of the uncertainty relations,
which are compatible with (\ref{emCR}).
This is discussed in a famous paper by Bohr and Rosenfeld (1950).

Second, the relation analogous to (\ref{continuity_eq}) does not hold
for the free photon field because the operator
\begin{equation}
\hat \jmath^{\nu} = \partial_{\mu} \hat F^{\mu\nu} =
\Box \hat A^{\nu} - \partial_{\mu} ( \partial^{\nu} \hat A^{\mu} ) = 0
\end{equation}
vanishes. This consequently means that there is no conservation
law for the number of photons. In other words the total number of photons
is uncertain. One therefore has to draw the conclusion that the concept of
a well-defined particle density (expressed by the number operator)
is not meaningful in the same way for the photon field as it is for
the electron field.
Or, in the words of Pauli (1933, p. 579):
{\em ''... da\ss \ f\"ur das Photonfeld ... der Begriff der
raum-zeitlich-lokalen Teilchendichte $W(\vec x,t)$ nicht sinnvoll existiert''
[ ... that for the photon field ... the notion of a particle density
$W(\vec x,t)$ located in space-time has no meaningful existence}
(translation by the author)].

\section{Multiple Quantization}

In the 1950s von~Weizs\"acker
(1955; 1958; von~Weizs\"acker, von~Weizs\"acker et al., 1958)
introduced both the idea of what was later called the quantum theory of
ur-alternatives (''ur-theory'') and his concept of multiple quantization.
Both ideas are related to each other.

\subsection{The Quantum Theory of Ur-Alternatives}

The ur-theory is a program to understand the unity of physics and is
based on the simplest possible object which can be found in quantum
theory: the quantized binary alternative (shortly ''ur-object'' or
''ur''). It is not the intention of this paper to describe the structure
of ur-theory (von~Weizs\"acker, 1985, Chapters 9 and 10);
only a short introduction to the basic idea shall be given.
In ur-theory the assumption is that the three-dimensionality
of position space is a consequence of the symmetry group of the ur,
which is essentially $SU(2)$.
Moreover, the homogeneous space of $SU(2)$, which is $\setS^3$,
can be looked upon as a model of our cosmos.
The argument for this is that if quantum theory gives the
fundamental structure of any physical theory,
then any physical object must be described by a Hilbert space
which in any case can be embedded into a tensor product space of urs.
Thus any physical object can be trivially build up from urs
and therefore the symmetry properties
of the position space have to be the symmetry properties of urs.
In ur-theory the line of argument is turned around:
the symmetry of position space is regarded as a consequence
of the symmetry of urs.
This points toward a close connection between empirical alternatives
and their testability in space (Lyre, 1995).

Thus an ur-alternative turns out to be the fundamental object in
physics. But in standard physics we deal with particles and
fields as described above.
This leads to the concept of multiple quantization.

\subsection{The Statistical Interpretation of Quantization}
\label{MQ_and_SI}

Let us now ask about the meaning of quantization and suppose quantum theory
to be fundamental. Thus we do not want to introduce quantum theory from
classical mechanics via the ''correspondence principle''.
Instead we will follow von Weizs\"acker's proposal of the connection
between quantum theory and probability theory.

Let us call the n possible answers, excluding each other,
\begin{equation}\label{n_alternative}
a_k \quad (k=1 ... n)
\end{equation}
to a given question an {\em n-fold-alternative}.
Then the complex numbers
\begin{equation}\label{truth}
\psi_k \quad (k=1 ... n)
\end{equation}
should be the corresponding truth values.
If $\psi$ is normalized, then
\begin{equation}
p_k=\psi_k^* \psi_k
\end{equation}
is the probability to find $a_k$.
Now probability can be defined as the
{\em expectation value of a relative frequency} $f_k = \frac{n_k}{n}$
(Drieschner, 1979)
\begin{equation}\label{probability}
p_k = E\left(f_k\right) = \sum_{f_k} p\left(f_k\right) \ f_k .
\end{equation}
This definition fits the fact that in real measurements only the
number $n_k$ of the occurrences of $a_k$ in a series of $n$
experiments is observed. Therefore in quantum theory $n_k$
has to be regarded as an operator. It turns out that
\begin{equation}
\hat n_k = \hat \psi_k^+ \hat \psi_k
\end{equation}
is a suitable choise, whereas the new opeators $\hat \psi_k^+$,
$\hat \psi_k$ obey certain commutation relations and act as creation
and annihilation operators of states $\psi_k$.

From
\begin{equation}
\left\langle \hat\psi \left| \hat\psi \right\rangle \right. =
\sum_k \hat \psi_k^+ \hat \psi_k
= \sum_k \hat n_k = \hat n
\end{equation}
it follows that
{\em the operator $\hat \psi$ of the next level of quantization must be
interpreted as a totality of $n$ objects of the level below - each one
described by a single wave function $\psi$.}
One therefore is led to a statistical interpretation of the
quantization procedure.

The definition (\ref{probability}) has yet another consequence.
On the first view it looks like a circular definition:
the probability $p_k$ is defined by another probability $p(f_k)$.
But one has to keep in mind that $p\left(f_k\right)$
is a probability of the next-higher level.
It describes the probability to find a series of experiments
(where $a_k$ was found with the relative frequency $f_k$)
in a series of series of experiments.
This new probability again refers to a probability of a higher level
and so on. Thus the step-like structure of probability theory appears
and, because of the connection between quantization and probability as
described above, this leads - by the same argument - to a
step-like structure of quantization. Therefore there should be not only
two, but multiple levels of quantization (von~Weizs\"acker, 1973).

\subsection{Multiple Quantization in Ur-Theory}

In ur-theory one starts with a simple, empirically testable, binary
alternative $a_r \ (r=1,2)$.
The first quantization of $a_r$ leads to the complex spinor $u_r$.
On the second level of quantization one has to
introduce the ur-operators $\hat u_r^+$, $\hat u_r$.
It was found that the momentum states of massless and massive particles
can be build up from creation and annihilation operators of urs and
anti-urs $(r=1,...4)$.
The appropriate commutation relations for these operators represent a
parabose-statistics of urs (G\"ornitz et al., 1992).
Thus the quantum field theory of particles such as quarks and leptons
(see Section \ref{dirac_field}) appears - in the light of ur-theory - as
the third level of quantization of the alternative $a_r$.
In view of the problems with the statistical interpretation
of the quantized electromagnetic field (see Section \ref{EM_field}),
the question arises of whether the photon should be build up from urs
in the same way as particles, or, if not, in what other way?
Surely the gauge theoretic character of the interaction fields
must be explained in ur-theory, but this leads to open questions
concerning the problem of interaction in ur-theory which will
not be discussed in this paper.

\section{A General Theory of Information}

From the interpretational point of view the theory of ur-objects
must be regarded as a {\em quantum theory of information}
consequently thought to its end.
An ur-alternative represents exactly
{\em one bit of potential information}.
The question now is: what is the meaning of the different levels of
quantization within the framework of a quantum theory of information?

For this purpose one has to be aware of the difference between
{\em syntactic} and {\em semantic information.}
I call syntactic information an
{\em amount of structural distinguishability}
which can be measured in $bits$.
Beyond this the semantic aspect of information takes care of the fact
that information only exists under a certain concept or
on a certain semantic level.
For example, a letter printed on a paper refers to different amounts
of information if it is regarded under the concept ''letter of an
alphabet of a certain language'' or under the concept ''molecules of
printer's ink''.
The statistical interpretation of quantization stresses the
importance of the forming of collectives, i.e., a wave function of a
certain level of quantization describes a totality of objects of the
level below. This is in a certain way analogous to the forming of
concepts, e.g., the concept ''animal'' describes the totality of cats,
dogs, snakes, elephants, mosquitoes and so on.

In the light of multiple quantization in ur-theory we get the following
semantic levels:
An ur-object represents the simplest structural distinction which can
be made in empirical science: a spatial yes-no alternative - one bit
of information. The next level of quantization refers to particles.
The concept ''particle'' describes a totality of urs which are to be
regarded as the field quanta of a particle.
At the next level , the level of quantum field theory,
the objects of the level below, i.e., particles, become field quanta
for themselves, i.e., the former ''concepts'' must now be regarded as
''syntactic elements'' under the new concept
of the quantized particle field.

This is exactly my concluding assumption:
quantum theory must be regarded as a general theory of information
and quantization has to be understood as the forming of concepts or
semantic levels which are necessary for
the existence of information in general.
In ur-theory the problem still remains
of what status the interaction fields
in this information-theoretic view will have.

\subsection*{Acknowledgments}

I thank Prof.~M.~Drieschner for his support and
St.~Kretzer for helpful remarks on the manuscript.
I also thank Prof.~C.~F.~von~Weizs\"acker for many stimulating discussions.
I am grateful to the Graduiertenf\"orderung of the Ruhr-University Bochum
for financial support.

\section*{References}

\begin{description}

\item[{\normalsize Bohr, N. and Rosenfeld, L. (1950).}]
Field and Charge Measurements in Quantum Electrodynamics,
{\em Physical Review}, 78(6):794.

\item[{\normalsize Drieschner, M. (1979).}]
{\em Voraussage - Wahrscheinlichkeit - Objekt. \"Uber die
  begrifflichen Grundlagen der Quantenmechanik},
Lecture Notes in Physics 99, Springer, Berlin.

\item[{\normalsize
G\"ornitz, Th., Graudenz, D., and von~Weizs\"acker, C. F. (1992).}]
Quantum Field Theory of Binary Alternatives,
{\em International Journal of Theoretical Physics}, 31(11):1929-1959.

\item[{\normalsize Lyre, H. (1995).}]
The Quantum Theory of Ur Objects as a Theory of
  {I}nformation,
{\em International Journal of Theoretical Physics}, 34(8):1541-1552.

\item[{\normalsize Pauli, W. (1933).}]
Einige die Quantenmechanik betreffende Erkundigungsfragen,
{\em Zeitschrift f\"ur Physik}, 80:573-586.

\item[{\normalsize von~Weizs\"acker, C. F. (1955).}]
Komplementarit\"at und Logik,
{\em Die Naturwissenschaften}, 42:521-529, 545-555.

\item[{\normalsize von~Weizs\"acker, C. F. (1958).}]
Die Quantentheorie der einfachen Alternative
          (Komplementarit\"at und Logik II),
{\em Zeitschrift f\"ur Naturforschung}, 13a:245-253.

\item[{\normalsize von~Weizs\"acker, C. F. (1973).}]
Probability and Quantum Mechanics,
{\em British Journal for the Philosophy of Science}, 24:321--337.

\item[{\normalsize von~Weizs\"acker, C. F. (1985).}]
{\em Aufbau der Physik},
Hanser, Munich.

\item[{\normalsize von~Weizs\"acker, C. F., Scheibe, E.,
      and S\"ussmann, G., (1958).}]
Komplementarit\"at und Logik III. Mehrfache Quantelung.
{\em Zeitschrift f\"ur Naturforschung}, 13a:705.

\end{description}

\end{document}